\documentclass[12pt]{elsarticle}
\usepackage{amssymb}
\usepackage{url}

\newenvironment{myenumerate}{
\begin{enumerate}
  \setlength{\itemsep}{1pt}
  \setlength{\parskip}{0pt}
  \setlength{\parsep}{0pt}}{\end{enumerate}
}

\newenvironment{myitemize}{
\begin{itemize}
  \setlength{\itemsep}{1pt}
  \setlength{\parskip}{0pt}
  \setlength{\parsep}{0pt}}{\end{itemize}
}

\begin{document}

\begin{frontmatter}

\setcounter{page}{1}

\title{Numerical Evaluation of Algorithmic Complexity for Short Strings: A Glance Into the Innermost Structure of Randomness}
\author{Jean-Paul Delahaye\fnref{label1}}
\ead{delahaye@lifl.fr}
\fntext[label1]{Laboratoire d'Informatique Fondamentale de Lille (CNRS), Universit\'e de Lille 1.}
\author{Hector Zenil\fnref{label2}}
\ead{hectorz@labores.eu}

\maketitle

\begin{abstract}
We describe an alternative method (to compression) that combines several theoretical and experimental results to numerically approximate the algorithmic (Kolmogorov-Chaitin) complexity of all $\sum_{n=1}^82^n$ bit strings up to 8 bits long, and for some between 9 and 16 bits long. This is done by an exhaustive execution of all deterministic 2-symbol Turing machines with up to 4 states for which the halting times are known thanks to the Busy Beaver problem, that is 11\,019\,960\,576 machines. An output frequency distribution is then computed, from which the algorithmic probability is calculated and the algorithmic complexity evaluated by way of the (Levin-Zvonkin-Chaitin) coding theorem.
\end{abstract}


\end{frontmatter}

\section{Overview}
\label{overview}

The most common approach to calculate the algorithmic complexity of a string is the use of compression algorithms exploiting the regularities of the string and producing shorter compressed versions. The length of a compressed version of a string is an upper bound of the algorithmic complexity of the string $s$.

In practice, it is a known problem that one cannot compress short strings, shorter, for example, than the length in bits of the compression program which is added to the compressed version of $s$, making the result (the program producing $s$) sensitive to the compressor choice and the parameters involved. However, short strings are quite often the kind of data encountered in many practical settings. While compressors' asymptotic behavior guarantees the eventual convergence to the algorithmic complexity of $s$, thanks to the invariance theorem (to be enunciated later), measurements differ considerably in the domain of short strings. A few attempts to deal with this problem have been reported before \cite{speidel}. The conclusion is that estimators are always challenged by short strings.

Attempts to compute the uncomputable are always challenging, see for example \cite{rado,brady,marxen} and more recently \cite{caludeomega} and \cite{hertel}. This often requires combining theoretical and experimental results. In this paper we describe a method to compute the algorithmic complexity (hereafter denoted by $C(s)$) of (short) bit strings by running a set of (relatively) large number of Turing machines for which the halting runtimes are known thanks to the Busy Beaver problem \cite{rado}. 

In the spirit of the experimental paradigm suggested in \cite{wolfram}, the method in this paper describes a way to find the shortest program given a standard formalism of Turing machines, executing all machines from the shortest (in number of states) to a certain (small) size one by one recording how many of them produce a string and then using a theoretical result linking this string frequency with the algorithmic complexity of a string.

The result is a novel approach that we put forward for numerically calculate the complexity of short strings as an alternative to the indirect method using compression algorithms. The procedure makes use of a combination of results from related areas of computation, such as the concept of halting probability \cite{chaitin}, the Busy Beaver problem \cite{rado}, algorithmic probability \cite{solomonoff}, Levin's semi-measure and Levin-Zvonkin-Chaitin's coding theorem (from now on \emph{coding theorem}) \cite{levin,levin2}.

The approach, never attempted before to the authors' knowledge, consists in the thorough execution of all 2-symbol Turing machines up to 4 states (the exact model is described in \ref{D}) which, upon halting, generate a set of output strings from which a frequency distribution is calculated to obtain the algorithmic probability of a string. The algorithmic complexity of a string can then be evaluated from the algorithmic probability using the coding theorem. 

The paper is structured as follows. In section \ref{preliminaries} it is introduced the various theoretical concepts and experimental results utilized in the experiment, providing essential definitions and referring the reader to the relevant papers and textbooks. Section \ref{D} introduces the definition of our empirical probability distribution $D$. In \ref{method} we present the methodology for calculating $D$. In \ref{results} we calculate $D$ and provide numerical values of the algorithmic complexity for short strings by way of the theory presented in \ref{preliminaries}, particularly the coding theorem. Finally, in \ref{conclusions} we summarize, discuss possible applications, and suggest potential directions for further research.

\section{Preliminaries}
\label{preliminaries}

\subsection{The Halting problem and Chaitin's $\Omega$}

As widely known, the Halting problem for Turing machines is the problem of deciding whether an arbitrary Turing machine $T$ eventually halts on an arbitrary input $s$. Halting computations can be recognized by simply running them for the time they take to halt. The problem is to detect non-halting programs, about which one cannot know if the program will run forever or will eventually halt. An elegant and concise representation of the halting problem is Chaitin's irrational number $\Omega$ \cite{chaitin}, defined as the halting probability of a universal computer programmed by coin tossing. Formally,\\

\noindent\textsc{Definition 1.} $0 < \Omega = \sum_{\normalsize{p\textbf{ }halts}} 2^{-|p|} < 1$
\noindent with $|p|$ the size of $p$ in bits.\\

$\Omega$ is the halting probability of a universal (prefix-free\footnote{A set of programs A is prefix-free if there are no two programs $p_1$ and $p_2$ such that $p_2$ is a proper extension of $p_1$. Kraft's inequality \cite{calude} guarantees that for any prefix-free set A, $\sum_{x\in A} 2^{-|x|} \leq 1$.}) Turing machine running a random program (a sequence of fair coin flip bits taken as a program).

For an $\Omega$ number one cannot compute more than a finite number of digits. The numerical value of $\Omega = \Omega_U$ depends on the choice of universal Turing machine $U$. There are, for example, $\Omega$ numbers for which no digit can be computed \cite{solovay}.

Knowing the first $n$ bits of an $\Omega$ allows to determine whether a program of length $\leq n$ bits halts by simply running all programs in parallel until the sum exceeds that $\Omega$. All programs with length $\leq n$ not halting yet will never halt. Using these kind of arguments, Calude and Stay \cite{calude2} have shown that most programs either stop ``quickly'' or never halt because the halting runtime (and therefore the length of the output upon halting) is ultimately bounded by its program-size complexity. The results herein connect theory with experiments by providing empirical values of halting times and string length frequencies.

\subsection{Algorithmic (prefix-free) complexity}

The algorithmic complexity $C_U(s)$ of a string $s$ with respect to a universal Turing machine $U$, measured in bits, is defined as the length in bits of the shortest (prefix-free) Turing machine $U$ that produces the string $s$ and halts \cite{solomonoff,kolmogorov,levin,chaitin}. Formally,\\

\noindent\textsc{Definition 2.} $C_U(s) = \min\{|p|, U(p)=s\}$
\noindent where $|p|$ is the length of $p$ measured in bits.\\

This complexity measure clearly seems to depend on $U$, and one may ask whether there exists a Turing machine which yields different values of $C(s)$. The answer is that there is no such Turing machine which can be used to decide whether a short description of a string is the shortest (for formal proofs see  \cite{calude,li}).

The ability of universal machines to efficiently simulate each other implies a corresponding degree of robustness. The invariance theorem \cite{solomonoff} states that if $C_U(s)$ and $C_{U^\prime}(s)$ are the shortest programs generating $s$ using the universal Turing machines $U$ and $U^\prime$ respectively, their difference will be bounded by an additive constant independent of $s$. Formally: \\

\noindent\textsc{Theorem (invariance \cite{solomonoff}) 1.} $|C_U(s) - C_{U^\prime}(s)| \leq c_{_{U,U^\prime}}$\\

A major drawback of $C$ as a function taking $s$ to the length of the shortest program producing $s$, is its non-computability proven by reduction to the halting problem. In other words, there is no program which takes a string $s$ as input and produces the integer $C(s)$ as output.

\subsection{Algorithmic probability}

Deeply connected to Chaitin's halting probability $\Omega$, is Solomonoff's concept of algorithmic probability, independently proposed and further formalized by Levin's \cite{levin} semi-measure herein denoted by $m(s)$.

Unlike Chaitin's $\Omega$, it is not only whether a program halts or not that matters for the concept of algorithmic probability; the output and halting time of a halting Turing machine are also relevant in this case.

Levin's semi-measure $m(s)$ is the probability of producing a string $s$ with a random program $p$ (i.e. every bit of $p$ is the result of an independent toss of a fair coin) when running on a universal prefix-free Turing machine $U$. Formally,\\

\noindent\textsc{Definition 3.} $m(s) = \sum_{p : U(p) = s} 2^{-|p|}$\\

Levin's probability measure induces a distribution over programs producing $s$, assigning to the shortest program the highest probability and smaller probabilities to longer programs.

There is a theorem connecting algorithmic probability to algorithmic complexity. Algorithmic probability is related to algorithmic complexity in that $m(s)$ is at least the maximum term in the summation of programs given that it is the shortest program that has the greater weight in the summation of the fractions defining $m(s)$. Formally, the theorem states that the following relation holds:\\

\noindent\textsc{Theorem (coding theorem \cite{calude}) 2.} $-\log_2\normalsize{ }m(s) = C(s) + O(1)$\\

Nevertheless, $m(s)$ as a function of $s$ is, like $C(s)$ and Chaitin's $\Omega$, noncomputable due to the halting problem\footnote{An important property of $m$ as semi-measure is that it dominates any other effective semi-measure $\mu$ because there is a constant $c_\mu$ such that, for all $s$, $m(s) \geq c_\mu\mu(s)$. For this reason $m(s)$ is often called a \emph{universal distribution} \cite{kirchherr}.}.

\subsection{The Busy Beaver problem: Solving the halting problem for small Turing machines}
\label{beaver}

\noindent\textsc{Notation:} We denote by $(n,2)$ the class (or space) of all $n$-state 2-symbol Turing machines (with the halting state not included among the $n$ states).\\

\noindent\textsc{Definition 4.} \cite{rado} If $\sigma_T$ is the number of 1s on the tape of a Turing machine $T$ upon halting, then:  $\sum(n)=\max{\{\sigma_T : T\in(n,2) \normalsize{\textbf{ }T(n)\textbf{ }halts}\}}$.\\

\noindent\textsc{Definition 5.} \cite{rado} If $t_T$ is the number of steps that a machine $T$ takes upon halting, then $S(n)=\max{\{t_T : T\in(n,2) \normalsize{\textbf{ }T(n)\textbf{ }halts}\}}$.\\

$\sum(n)$ and $S(n)$ as defined (and denoted by Busy Beaver functions) in 4 and 5 are noncomputable by reduction to the halting problem \cite{rado}. Yet values are known for $(n,2)$ with $n \leq 4$. The solution for $(n,2)$ with $n<3$ is trivial, the process leading to the solution in $(3,2)$ is discussed by Lin and Rado \cite{lin}, and the process leading to the solution in $(4,2)$ is discussed in \cite{brady}.

A program showing the evolution of all known Busy Beaver machines developed by one of this paper's authors is available online \cite{busyhz}. The Turing machine model followed in this paper is the same as the one described for the Busy Beaver problem as introduced by Rado \cite{rado} .

\section{The empirical distribution $D$}
\label{D}

It is important to describe the Turing machine formalism because exact values of algorithmic probability for short strings will be provided under this chosen standard model of Turing machines.

\noindent\textsc{Definition 6.} Consider a Turing machine with the binary alphabet $\Sigma=\{0,1\}$ and $n$ states $\{1,2, \ldots n\}$ and an additional Halt state denoted by 0 (just as defined in Rado's original Busy Beaver paper \cite{rado}).\\

The machine runs on a $2$-way unbounded tape. At each step: 

\begin{myenumerate}
\item the machine's current ``state'' (instruction); and
\item the tape symbol the machine's head is scanning
\end{myenumerate}

define each of the following: 

\begin{myenumerate}
\item a unique symbol to write (the machine can overwrite a $1$ on a $0$, a $0$ on a $1$, a $1$ on a $1$, and a $0$ on a $0$);
\item a direction to move in: $-1$ (left), $1$ (right) or $0$ (none, when halting); and
\item a state to transition into (may be the same as the one it was in).
\end{myenumerate}

The machine halts if and when it reaches the special halt state 0. There are  $(4n + 2)^{2n}$ Turing machines with $n$ states and 2 symbols according to the formalism described above.

No transition starting from the halting state exists, and the blank symbol is one of the 2 symbols (0 or 1) in the first run, while the other is used in the second run (in order to avoid any asymmetries due to the choice of a single blank symbol). In other words, we run each machine twice, one with 0 as the blank symbol (the symbol with which the tape starts out and is filled with), and an additional run with 1 as the blank symbol\footnote{Due to the symmetry of the computation, there is no real need to run each machine twice; one can \emph{complete} the string frequencies assuming that each string produced its reversed and complemented version with the same frequency, and then group and divide by symmetric groups. A more detailed explanation of how this is done is in  \cite{delahayezenil}.}. The output string is taken from the number of contiguous cells on the tape the head of the halting $n$-state machine has gone through. A machine produces a string upon halting.\\

\noindent\textsc{Definition 7.} $D(n)$ is the function that retrieves the number of machines that halt (denoted by $d(n)$) in $(n,2)$ and then assigns to every string $s$ produced by $(n,2)$ the quotient: (number of times that a machine in $(n,2)$ produces $s$) / (number of machines in $(n,2)$ that halt).\\

Examples of $D(n)$ for $n=1,n=2$:

\begin{center}
$d(1)=24$, $D(1) = 0 \rightarrow 0.5; 1 \rightarrow 0.5$\\
$d(2)=6088$, $D(2) = 0 \rightarrow 0.328; 1 \rightarrow 0.328; 00 \rightarrow .0834 \ldots$\\
\end{center}

Tables 1, 2 and 3 in \ref{results} show the results for $D(1)$, $D(2)$ and $D(3)$, and Table 4 the top ranking of $D(4)$.\\

\noindent\textsc{Theorem 3.} $D(n)$ is noncomputable.\\

\noindent\textsc{Proof (by reduction to the halting problem):} The result is obvious, since from the knowledge of the number of $n$-state Turing machines that halt, it is easy to know for every Turing machine if it stops or not by the following argument (by contradiction): Assume $D(n)$ is computable. Let $T$ be any arbitrary Turing machine. To solve the halting problem for $T$, calculate $D(n)$, where $n$ is the number of states in $T$. Suppose that (by hypothesis) $D(n)$ outputs $d(n)$ and the assignation list of strings and frequencies. Run all possible $n$-state Turing machines in parallel, and wait until $d(n)$ many of the machines have halted.  If $T$ is one of the machines that has halted, then $T$ halts. Otherwise, $T$ doesn't halt.  We have just shown that if $D(n)$ were computable, then the halting problem would be solvable. Since the halting problem is known to be unsolvable, $D$ must be noncomputable.

Exact values of $D(n)$ can be, however, calculated for small Turing machines because of the known values (in particular $S(n)$) of the Busy Beaver problem for $n<5$. For example, for $n=4$, $S(4)=107$, so we know that any machine running more than 107 steps will never halt and so we stop it thereafter.

For each Busy Beaver candidate with $n>4$ states, a sample of Turing machines running up to the candidate $S(n)$ is also possible. As for Rado's Busy Beaver functions $\sum(n)$ and $S(n)$, $D(n)$ is also approachable \emph{from above}. For larger $n$, sampling methods asymptotically converging to $D(n)$ can be used to approximate $D(n)$. In section \ref{results} we provide exact values of $D(n)$ for $n<5$ thanks to the the Busy Beaver known values.

Another property shared between $D(n)$ and the Busy Beaver problem is that $D(4)$, just as the values of the Busy Beaver, is well-defined in the sense that the calculation of the digits of $D(n)$ are fully determined once calculated, but the calculation of $D(n)$ rapidly becomes impractical to calculate, for even a slightly larger number of states. Our quest is thus similar in several respects to the Busy Beaver problem or the calculation of the digits of Chaitin's $\Omega$ number. The main underlying difficulty in analyzing thoroughly a given class of machines is the undecidability of the halting problem, and hence the uncomputability of the related functions.

\section{Methodology}
\label{method}

The approach for evaluating the complexity $C(s)$ of a string $s$ presented herein is limited by (1) the halting problem and (2) computing time constraints. Restriction (1) was overcome using the values of the Busy Beaver problem providing the halting times for all Turing machines starting with a blank tape. Restriction (2) represented a challenge in terms of computing time and programming skills. It is also the same restriction that has kept others from attempting to solve the Busy Beaver problem for a greater number of states.  We were able to compute up to about $1.3775 \times 10^9$ machines per day or 15\,943 per second, taking us about 9 days\footnote{Running on a MacBook Intel Core Duo at 1.83Ghz with 2Gb. of RAM memory and a solid state hard drive, using the TuringMachine[] function available in $Mathematica$ 8 for $n<4$ and a C++ program for $n=4$. Since for $n=4$ there were  $2.56 \times 10^8$ machines involved, running on both 0 and 1 as blank, further optimizations were required. The use of a Bignum library and an actual enumeration of the machines rather than producing the rules beforehand (which would have meant overloading the memory even before the actual calculation) was necessary.} to run all $(4,2)$ Turing machines each up to the number of steps bounded by the Busy Beaver values.

Just as it is done for solving small values of the Busy Beaver problem, we rely on the experimental approach to analyze and describe a computable fraction of the uncomputable. A similar quest for the calculation of the digits of a Chaitin's $\Omega$ number was undertaken by Calude et al. \cite{caludeomega}, but unlike Chaitin's $\Omega$, the calculation of $D(n)$ does not depend on the enumeration of Turing machines (because ). It is easy to see that every $(2,n)$ Turing machine contributing to $D(n)$ is included in $D(n+1)$ simply because every Turing machine in $(2,n)$ is also in $(2,n+1)$.

\subsection{Numerical calculation of $D$}
\label{calculating}

We consider the space $(n,2)$ of Turing machines with $0<n<5$. The halting ``history'' and output probability followed by their respective runtimes, presented in Tables 1, 2 and 3, show the times at which the programs in the domain of M halt, the frequency of the strings produced, and the time at which they halted after writing down the output string on their tape.

We provide exact values for $n = \{2, 3, 4\}$ in the Results \ref{results}. We derive $D(n)$ for $n<5$ from counting the number of n-strings produced by all $(n,2)$ Turing machines upon halting. We define $D$ to be an $empirical$ \emph{universal distribution} in  Levin's  sense, and calculate the algorithmic complexity $C$ of a string $s$ in terms of $D$ using the coding theorem, from which we won't escape to an additive constant introduced by the application of the coding theorem, but the additive constant is common to all values and therefore should not impact the relative order. One has to bear in mind, however, that the tables in section \ref{results} should be read as dependent of this last-step additive constant because using the coding theorem as an approximation method fixes a prefix-free universal Turing machine via that constant, but according to the choices we make this seems to be the most natural way to do so as an alternative to other indirect choosing procedures.

We calculated the 72, 20\,000, 15\,059\,072 and 22\,039\,921\,152 two-way tape Turing machines started with a tape filled with 0s and 1s for $D(2)$, $D(3)$ and $D(4)$\footnote{The space occupied by the outputs building $D(4)$ was 77.06Gb.}. The number of Turing machines to calculate grows exponentially with the number of states. For $D(5)$ there are 53\,119\,845\,582\,848 machines to calculate, which makes the task as difficult as finding the Busy Beaver values for $\sum(5)$ and $S(5)$, Busy Beaver values which are currently unknown but for which the best candidate may be $S(5)=47\,176\,870$ which makes the exploration of $(5,2)$ a greatest challenge.

Although several ideas exploiting symmetries to reduce the total number of Turing machines have been proposed and used for finding Busy Beaver candidates \cite{brady,machlin,holkner} in large spaces such as $n \geq 5$, to preserve the structure of the data we couldn't apply all of them. This is because, unlike the Busy Beaver challenge, in which only the maximum values are important, the construction of a probability distribution requires every output to be equally considered. Some reduction techniques were, however, utilized, such as running only one-direction rules with a tape only filled with 0s and then completing the strings by reversion and complementation to avoid running every machine a second time with a tape filled with 1s. For an explanation of how we counted the number of symmetries to recuperate the outputs of the machines that were skipped see \cite{delahayezenil}.

\section{Results}
\label{results}

\subsection{Algorithmic probability tables}

$D(1)$ is trivial. $(1,2)$ Turing machines produce only two strings, with the same number of machines producing each. The Busy Beaver values for $n=1$ are $\sum(1)=1$ and $S(1)=1$. That is, all machines that halt do so after 1 step, and print at most one symbol.

\begin{table}[!htdp]
\label{firsttable}
\caption{Distribution ($D(1)$) from the $d(1)=24$ machines in $(1,2)$ that halt, out of a total of 64 Turing machines.}
\begin{center}
\begin{tabular}{|c|c|c|} 
  \hline
   0 : 0.5\\
   1 : 0.5\\
  \hline
\end{tabular}
\end{center}
\label{default}
\end{table}

The Busy Beaver values for $n=2$ are $\sum(1)=4$ and $S(1)=6$. $D(2)$ is quite simple but starts to display some basic structure, such as a clear correlation between string length and occurrence, following what may be an exponential decrease in the number of string occurrences:

\begin{center}
$P(|s|=1)=0.657$\\
$P(|s|=2)=0.333$\\
$P(|s|=3)=0.0065$\\
$P(|s|=4)=0.0026$\\
\end{center}

\begin{table}[!htdp]
\caption{Distribution $D(2)$ from 6\,088 $(2,2)$ out of 20\,000 Turing machines that halt. Each string is followed by its probability (from the number of times produced), sorted from highest to lowest.}
\begin{center}
\begin{tabular}{|l|l|}
  \hline
0 : .328 & 010 : .00065\\
1 : .328 & 101 : .00065\\
00 : .0834 & 111 : .00065\\
01 : .0834 & 0000 : .00032\\
10 : .0834 &0010 : .00032\\
11 : .0834 & 0100 : .00032\\
001 : .00098 & 0110 : .00032\\
011 : .00098 &1001 : .00032\\
100 : .00098 & 1011 : .00032\\
110 : .00098 &1101 : .00032\\
000 : .00065& 1111 : .00032\\
  \hline
\end{tabular}
\end{center}
\label{default}
\end{table}

\noindent Among the various facts one can draw from $D(2)$, there are:

\begin{myitemize}
\item There are $d(2)=6088$ machines that halt out of the $20\,000$ Turing machines in $(2,2)$ as the result of running every machine over a tape filled with 0 and then again over a tape filled with 1.
\item The relative string order in $D(1)$ is preserved in $D(2)$.
\item A fraction of $1/3$ of the total machines halt while the remaining $2/3$ do not. That is, 24 among 72 (running each machine twice with tape filled with 1 and 0 as explained before).
\item The longest string produced by $D(2)$ is of length 4.
\item $D(2)$ does not produce all $\sum_1^4 2^n=30$ strings shorter than 5, only 22. The missing strings are 0001, 0101 and 0011 never produced, hence neither were their complements and reversions: 0111, 1000, 1110, 1010 and 1100. 
\end{myitemize}

\begin{table}[htdp]\renewcommand{\arraystretch}{.73}\addtolength{\tabcolsep}{-1pt}
\caption{Probability distribution ($D(3)$) produced by all the 15\,059\,072 Turing machines in $(3,2)$.}
\begin{center}
\begin{tabular}{|l|l|l|} 
  \hline
\footnotesize 0 : 0.250 & \footnotesize 11110 : 0.0000470 & \footnotesize 100101 : \footnotesize{1.43$\times 10^{-6}$} \\
\footnotesize 1 : 0.250 & \footnotesize 00100 : 0.0000456 & \footnotesize 101001 : \footnotesize{1.43$\times 10^{-6}$} \\
\footnotesize 00 : 0.101 & \footnotesize 11011 : 0.0000456 & \footnotesize 000011 : \footnotesize{9.313$\times 10^{-7}$} \\
\footnotesize 01 : 0.101 & \footnotesize 01010 : 0.0000419 & \footnotesize 000110 : \footnotesize{9.313$\times 10^{-7}$} \\
\footnotesize 10 : 0.101 & \footnotesize 10101 : 0.0000419 & \footnotesize 001100 : \footnotesize{9.313$\times 10^{-7}$} \\
\footnotesize 11 : 0.101 & \footnotesize 01001 : 0.0000391 & \footnotesize 001101 : \footnotesize{9.313$\times 10^{-7}$} \\
\footnotesize 000 : 0.0112 & \footnotesize 01101 : 0.0000391 & \footnotesize 001111 : \footnotesize{9.313$\times 10^{-7}$} \\
\footnotesize 111 : 0.0112 & \footnotesize 10010 : 0.0000391 & \footnotesize 010001 : \footnotesize{9.313$\times 10^{-7}$} \\
\footnotesize 001 : 0.0108 & \footnotesize 10110 : 0.0000391 & \footnotesize 010010 : \footnotesize{9.313$\times 10^{-7}$} \\
\footnotesize 011 : 0.0108 & \footnotesize 01110 : 0.0000289 & \footnotesize 010011 : \footnotesize{9.313$\times 10^{-7}$} \\
\footnotesize 100 : 0.0108 & \footnotesize 10001 : 0.0000289 & \footnotesize 011000 : \footnotesize{9.313$\times 10^{-7}$} \\
\footnotesize 110 : 0.0108 & \footnotesize 00101 : 0.0000233 & \footnotesize 011101 : \footnotesize{9.313$\times 10^{-7}$} \\
\footnotesize 010 : 0.00997 & \footnotesize 01011 : 0.0000233 & \footnotesize 011110 : \footnotesize{9.313$\times 10^{-7}$} \\
\footnotesize 101 : 0.00997 & \footnotesize 10100 : 0.0000233 & \footnotesize 100001 : \footnotesize{9.313$\times 10^{-7}$} \\
\footnotesize 0000 : 0.000968 & \footnotesize 11010 : 0.0000233 & \footnotesize 100010 : \footnotesize{9.313$\times 10^{-7}$} \\
\footnotesize 1111 : 0.000968 & \footnotesize 00011 : 0.0000219 & \footnotesize 100111 : \footnotesize{9.313$\times 10^{-7}$} \\
\footnotesize 0010 : 0.000699 & \footnotesize 00111 : 0.0000219 & \footnotesize 101100 : \footnotesize{9.313$\times 10^{-7}$} \\
\footnotesize 0100 : 0.000699 & \footnotesize 11000 : 0.0000219 & \footnotesize 101101 : \footnotesize{9.313$\times 10^{-7}$} \\
\footnotesize 1011 : 0.000699 & \footnotesize 11100 : 0.0000219 & \footnotesize 101110 : \footnotesize{9.313$\times 10^{-7}$} \\
\footnotesize 1101 : 0.000699 & \footnotesize 000000 : \footnotesize{3.733$\times 10^{-6}$} & \footnotesize 110000 : \footnotesize{9.313$\times 10^{-7}$} \\
\footnotesize 0101 : 0.000651 & \footnotesize 111111 : \footnotesize{3.733$\times 10^{-6}$} & \footnotesize 110010 : \footnotesize{9.313$\times 10^{-7}$} \\
\footnotesize 1010 : 0.000651 & \footnotesize 000001 : \footnotesize{2.793$\times 10^{-6}$} & \footnotesize 110011 : \footnotesize{9.313$\times 10^{-7}$} \\
\footnotesize 0001 : 0.000527 & \footnotesize 011111 : \footnotesize{2.793$\times 10^{-6}$} & \footnotesize 111001 : \footnotesize{9.313$\times 10^{-7}$} \\
\footnotesize 0111 : 0.000527 & \footnotesize 100000 : \footnotesize{2.793$\times 10^{-6}$} & \footnotesize 111100 : \footnotesize{9.313$\times 10^{-7}$} \\
\footnotesize 1000 : 0.000527 & \footnotesize 111110 : \footnotesize{2.793$\times 10^{-6}$} & \footnotesize 0101010 : \footnotesize{9.313$\times 10^{-7}$} \\
\footnotesize 1110 : 0.000527 & \footnotesize 000100 : \footnotesize{2.333$\times 10^{-6}$} & \footnotesize 1010101 : \footnotesize{9.313$\times 10^{-7}$} \\
\footnotesize 0110 : 0.000510 & \footnotesize 001000 : \footnotesize{2.333$\times 10^{-6}$} & \footnotesize 001110 : \footnotesize{4.663$\times 10^{-7}$} \\
\footnotesize 1001 : 0.000510 & \footnotesize 110111 : \footnotesize{2.333$\times 10^{-6}$} & \footnotesize 011100 : \footnotesize{4.663$\times 10^{-7}$} \\
\footnotesize 0011 : 0.000321 & \footnotesize 111011 : \footnotesize{2.333$\times 10^{-6}$} & \footnotesize 100011 : \footnotesize{4.663$\times 10^{-7}$} \\
\footnotesize 1100 : 0.000321 & \footnotesize 000010 : \footnotesize{1.863$\times 10^{-6}$} & \footnotesize 110001 : \footnotesize{4.663$\times 10^{-7}$} \\
\footnotesize 00000 : 0.0000969 & \footnotesize 001001 : \footnotesize{1.863$\times 10^{-6}$} & \footnotesize 0000010 : \footnotesize{4.663$\times 10^{-7}$} \\
 \footnotesize 11111 : 0.0000969 & \footnotesize 001010 : \footnotesize{1.863$\times 10^{-6}$} & \footnotesize 0000110 : \footnotesize{4.663$\times 10^{-7}$} \\
\footnotesize 00110 : 0.0000512 & \footnotesize 010000 : \footnotesize{1.863$\times 10^{-6}$} & \footnotesize 0100000 : \footnotesize{4.663$\times 10^{-7}$} \\
\footnotesize 01100 : 0.0000512 & \footnotesize 010100 : \footnotesize{1.863$\times 10^{-6}$} & \footnotesize 0101110 : \footnotesize{4.663$\times 10^{-7}$} \\
\footnotesize 10011 : 0.0000512 & \footnotesize 011011 : \footnotesize{1.863$\times 10^{-6}$} & \footnotesize 0110000 : \footnotesize{4.663$\times 10^{-7}$} \\
\footnotesize 11001 : 0.0000512 & \footnotesize 100100 : \footnotesize{1.863$\times 10^{-6}$} & \footnotesize 0111010 : \footnotesize{4.663$\times 10^{-7}$} \\
\footnotesize 00010 : 0.0000489 & \footnotesize 101011 : \footnotesize{1.863$\times 10^{-6}$} & \footnotesize 1000101 : \footnotesize{4.663$\times 10^{-7}$} \\
\footnotesize 01000 : 0.0000489 & \footnotesize 101111 : \footnotesize{1.863$\times 10^{-6}$} & \footnotesize 1001111 : \footnotesize{4.663$\times 10^{-7}$} \\
\footnotesize 10111 : 0.0000489 & \footnotesize 110101 : \footnotesize{1.863$\times 10^{-6}$} & \footnotesize 1010001 : \footnotesize{4.663$\times 10^{-7}$} \\
\footnotesize 11101 : 0.0000489 & \footnotesize 110110 : \footnotesize{1.863$\times 10^{-6}$} & \footnotesize 1011111 : \footnotesize{4.663$\times 10^{-7}$} \\
\footnotesize 00001 : 0.0000470 & \footnotesize 111101 : \footnotesize{1.863$\times 10^{-6}$} & \footnotesize 1111001 : \footnotesize{4.663$\times 10^{-7}$} \\
\footnotesize 01111 : 0.0000470 & \footnotesize 010110 : \footnotesize{1.43$\times 10^{-6}$} & \footnotesize 1111101 : \footnotesize{4.663$\times 10^{-7}$} \\
\footnotesize 10000 : 0.0000470 & \footnotesize 011010 : \footnotesize{1.43$\times 10^{-6}$} & \footnotesize \footnotesize{}\\
 \hline
\end{tabular}
\end{center}
\label{default}
\end{table}

Given the number of machines to run, $D(3)$ constitutes the first non trivial probability distribution to calculate. The Busy Beaver values for $n=3$ are $\sum(3)=6$ and $S(3)=21$.\\

\noindent Among the various facts for $D(3)$:

\begin{myitemize}
\item There are $d(3)=4\,294\,368$ machines that halt among the 15\,059\,072 in $(3,2)$. That is a fraction of 0.2851.
\item The longest string produced in $(3,2)$ is of length 7.
\item $D(3)$ has not all $\sum_1^7 2^n=254$ strings shorter than 7 but 128 only, half of all the possible strings up to that length.
\item $D(3)$ preserves the string order of $D(2)$.
\end{myitemize}

$D(3)$ ratifies the tendency of classifying strings by length with exponentially decreasing values. The distribution comes sorted by length blocks from which one cannot easily say whether those at the bottom are more random-looking than those in the middle, but one can definitely say that the ones at the top, both for the entire distribution and by length block, are intuitively the simplest. Both $0^k$ and its reversed $1^k$ for $n\leq8$ are always at the top of each block, with 0 and 1 at the top of them all. There is a single exception in which strings were not sorted by length, this is the string group $0101010$ and $1010101$ that are found four places further away from their length block, which we take as a second indication of a complexity classification becoming more visible since these 2 strings correspond to what one would intuitively consider less random-looking because  they are easily described as the repetition of two bits.

$D(4)$ with 22\,039\,921\,152 machines to run was a true challenge, both in terms of programming specification and computational resources. The Busy Beaver values for $n=4$ are $\sum(3)=13$ and $S(n)=107$. Evidently every machine in $(n,2)$ for $n\leq4$ is in $(4,2)$ because a rule in $(n,2)$ with $n\leq4$ is a rule in $(4,2)$. The results are presented in \ref{d} and it is important to notice that the table presents the top of a much larger classification available online at \url{http://www.algorithmicnature.org} under the paper title as additional material. Hence, among the 129 there are supposed to be the strings with greatest structure. The reader can verify that the closer to the bottom the more random-looking.\\

\begin{table}[htdp]\renewcommand{\arraystretch}{.75}\addtolength{\tabcolsep}{-1pt}
\label{d}
\caption{The top 129 strings from $D(4)$ with highest probability (therefore with lowest random complexity) from 1832 different produced strings.}
\begin{center}
\begin{tabular}{|l|l|l|}
\hline
\footnotesize 0 : 0.205 &\footnotesize  01101 : 0.000145 &\footnotesize  110111 : 0.0000138 \\
\footnotesize 1 : 0.205 &\footnotesize  10010 : 0.000145 &\footnotesize  111011 : 0.0000138 \\
\footnotesize 00 : 0.102 &\footnotesize  10110 : 0.000145 &\footnotesize  001001 : 0.0000117 \\
\footnotesize 01 : 0.102 &\footnotesize  01010 : 0.000137 &\footnotesize  011011 : 0.0000117 \\
\footnotesize 10 : 0.102 &\footnotesize  10101 : 0.000137 &\footnotesize  100100 : 0.0000117 \\
\footnotesize 11 : 0.102 &\footnotesize  00110 : 0.000127 &\footnotesize  110110 : 0.0000117 \\
\footnotesize 000 : 0.0188 &\footnotesize  01100 : 0.000127 &\footnotesize  010001 : 0.0000109 \\
\footnotesize 111 : 0.0188 &\footnotesize  10011 : 0.000127 &\footnotesize  011101 : 0.0000109 \\
\footnotesize 001 : 0.0180 &\footnotesize  11001 : 0.000127 &\footnotesize  100010 : 0.0000109 \\
\footnotesize 011 : 0.0180 &\footnotesize  00101 : 0.000124 &\footnotesize  101110 : 0.0000109 \\
\footnotesize 100 : 0.0180 &\footnotesize  01011 : 0.000124 &\footnotesize  000011 : 0.0000108 \\
\footnotesize 110 : 0.0180 &\footnotesize  10100 : 0.000124 &\footnotesize  001111 : 0.0000108 \\
\footnotesize 010 : 0.0171 &\footnotesize  11010 : 0.000124 &\footnotesize  110000 : 0.0000108 \\
\footnotesize 101 : 0.0171 &\footnotesize  00011 : 0.000108 &\footnotesize  111100 : 0.0000108 \\
\footnotesize 0000 : 0.00250 &\footnotesize  00111 : 0.000108 &\footnotesize  000110 : 0.0000107 \\
\footnotesize 1111 : 0.00250 &\footnotesize  11000 : 0.000108 &\footnotesize  011000 : 0.0000107 \\
\footnotesize 0001 : 0.00193 &\footnotesize  11100 : 0.000108 &\footnotesize  100111 : 0.0000107 \\
\footnotesize 0111 : 0.00193 &\footnotesize  01110 : 0.0000928 &\footnotesize  111001 : 0.0000107 \\
\footnotesize 1000 : 0.00193 &\footnotesize  10001 : 0.0000928 &\footnotesize  001101 : 0.0000101 \\
\footnotesize 1110 : 0.00193 &\footnotesize  000000 : 0.0000351 &\footnotesize  010011 : 0.0000101 \\
\footnotesize 0101 : 0.00191 &\footnotesize  111111 : 0.0000351 &\footnotesize  101100 : 0.0000101 \\
\footnotesize 1010 : 0.00191 &\footnotesize  000001 : 0.0000195 &\footnotesize  110010 : 0.0000101 \\
\footnotesize 0010 : 0.00190 &\footnotesize  011111 : 0.0000195 &\footnotesize  001100 : \footnotesize{9.943$\times 10^{-6}$} \\
\footnotesize 0100 : 0.00190 &\footnotesize  100000 : 0.0000195 &\footnotesize  110011 : \footnotesize{9.943$\times 10^{-6}$} \\
\footnotesize 1011 : 0.00190 &\footnotesize  111110 : 0.0000195 &\footnotesize  011110 : \footnotesize{9.633$\times 10^{-6}$} \\
\footnotesize 1101 : 0.00190 &\footnotesize  000010 : 0.0000184 &\footnotesize  100001 : \footnotesize{9.633$\times 10^{-6}$} \\
\footnotesize 0110 : 0.00163 &\footnotesize  010000 : 0.0000184 &\footnotesize  011001 : \footnotesize{9.3$\times 10^{-6}$} \\
\footnotesize 1001 : 0.00163 &\footnotesize  101111 : 0.0000184 &\footnotesize  100110 : \footnotesize{9.3$\times 10^{-6}$} \\
\footnotesize 0011 : 0.00161 &\footnotesize  111101 : 0.0000184 &\footnotesize  000101 : \footnotesize{8.753$\times 10^{-6}$} \\
\footnotesize 1100 : 0.00161 &\footnotesize  010010 : 0.0000160 &\footnotesize  010111 : \footnotesize{8.753$\times 10^{-6}$} \\
\footnotesize 00000 : 0.000282 &\footnotesize  101101 : 0.0000160 &\footnotesize  101000 : \footnotesize{8.753$\times 10^{-6}$} \\
\footnotesize 11111 : 0.000282 &\footnotesize  010101 : 0.0000150 &\footnotesize  111010 : \footnotesize{8.753$\times 10^{-6}$} \\
\footnotesize 00001 : 0.000171 &\footnotesize  101010 : 0.0000150 &\footnotesize  001110 : \footnotesize{7.863$\times 10^{-6}$} \\
\footnotesize 01111 : 0.000171 &\footnotesize  010110 : 0.0000142 &\footnotesize  011100 : \footnotesize{7.863$\times 10^{-6}$} \\
\footnotesize 10000 : 0.000171 &\footnotesize  011010 : 0.0000142 &\footnotesize  100011 : \footnotesize{7.863$\times 10^{-6}$} \\
\footnotesize 11110 : 0.000171 &\footnotesize  100101 : 0.0000142 &\footnotesize  110001 : \footnotesize{7.863$\times 10^{-6}$} \\
\footnotesize 00010 : 0.000166 &\footnotesize  101001 : 0.0000142 &\footnotesize  001011 : \footnotesize{6.523$\times 10^{-6}$} \\
\footnotesize 01000 : 0.000166 &\footnotesize  001010 : 0.0000141 &\footnotesize  110100 : \footnotesize{6.523$\times 10^{-6}$} \\
\footnotesize 10111 : 0.000166 &\footnotesize  010100 : 0.0000141 &\footnotesize  000111 : \footnotesize{6.243$\times 10^{-6}$} \\
\footnotesize 11101 : 0.000166 &\footnotesize  101011 : 0.0000141 &\footnotesize  111000 : \footnotesize{6.243$\times 10^{-6}$} \\
\footnotesize 00100 : 0.000151 &\footnotesize  110101 : 0.0000141 &\footnotesize  0000000 : \footnotesize{3.723$\times 10^{-6}$} \\
\footnotesize 11011 : 0.000151 &\footnotesize  000100 : 0.0000138 &\footnotesize  1111111 : \footnotesize{3.723$\times 10^{-6}$} \\
\footnotesize 01001 : 0.000145 &\footnotesize  001000 : 0.0000138 &\footnotesize  0101010 : \footnotesize{2.393$\times 10^{-6}$}\\
 \hline
\end{tabular}
\end{center}
\label{default}
\end{table}

\noindent Among the various facts from these results: 

\begin{myitemize}
\item There are $d(4)=5\,970\,768\,960$ machines that halt in $(4,2)$. That is a fraction of 0.27.
\item A total number of 1824 strings were produced in $(4,2)$.
\item The longest string produced  is of length 16 (only 8 among all the $2^{16}$ possible were generated).
\item The Busy Beaver machines (writing more 1s than any other and halting) found in $(4,2)$ had very low probability among all the halting machines: $pr$(11111111111101)$=2.01 \times 10^{-9}$. Because of the reverted string (10111111111111), the total probability of finding a Busy Beaver in $(4,2)$ is therefore $4.02 \times 10^{-9}$ only (or twice that number if the complemented string with the maximum number of 0s is taken).
\item The longest strings in $(4,2)$ were in the string groups represented by the following strings: 1101010101010101, 1101010100010101, 101010101010
1011 and 1010100010101011, each with about 5.4447$\times 10^{-10}$ probability, i.e. an even smaller probability than for the Busy Beavers, and therefore the most random in the classification.
\item $(4,2)$ produces all strings up to length 8, then the number of strings larger than 8 rapidly decreases. The following are the number of strings by length $|\{s : |s|=l\}|$ generated and represented in $D(4)$ from a total of 1824 different strings. From $i=1,\ldots,15$ the values l of $|\{s : |s|=n\}|$ are 2, 4, 8, 16, 32, 64, 128, 256, 486, 410, 252, 112, 46, 8, and 0, which indicated all $2^l$ strings where generated for $n\leq 8$.
\item While the probability of producing a string with an odd number of 1s is the same than the probability of producing a string with an even number of 1s (and therefore the same for 0s), the probability of producing a string of odd length is .559 and .441 for even length.
\item As in $D(3)$, where we report that one string group (0101010 and its reversion), in $D(4)$ 399 strings climbed  to the top and were not sorted among their length groups. 
\item In $D(4)$ string length was no longer a determinant for string positions. For example, between positions 780 and 790, string lengths are: 11, 10, 10, 11, 9, 10, 9, 9, 9, 10 and 9 bits.
\item $D(4)$ preserves the string order of $D(3)$ except in 17 places out of 128 strings in $D(3)$ ordered from highest to lowest string frequency. The maximum rank distance among the farthest two differing elements in $D(3)$ and $D(4)$ was 20, with an average of 11.23 among the 17 misplaced cases and a standard deviation of about 5 places. The Spearman's rank correlation coefficient between the two rankings had a critical value of 0.98, meaning that the order of the 128 elements in $D(3)$ compared to their order in $D(4)$ were in an interval confidence of high significance with almost null probability to have produced by chance.
\end{myitemize}

\begin{table}[htdp]
\caption{Probabilities of finding $n$ 1s (or 0s) in $(4,2)$.}
\begin{center}
\begin{tabular}{|r|r|}
\hline
 \normalsize{number} & \\
 \normalsize{$n$ of 1s  } & \normalsize{$pr(n)$ } \\
\hline
 1 & 0.472 \\
 2 & 0.167 \\
 3 & 0.0279 \\
 4 & 0.00352 \\
 5 & 0.000407 \\
 6 & 0.0000508 \\
 7 & \normalsize{6.5$\times 10^{-6}$} \\
 8 & \normalsize{1.31$\times 10^{-6}$} \\
 9 & \normalsize{2.25$\times 10^{-7}$} \\
 10 & \normalsize{3.62$\times 10^{-8}$} \\
 11 & \normalsize{1.61$\times 10^{-8}$} \\
 12 & \normalsize{1.00$\times 10^{-8}$} \\
 13 & \normalsize{4.02$\times 10^{-9}$}\\
 \hline
\end{tabular}
\end{center}
\label{default}
\end{table}

\begin{table}[htdp]
\caption{String groups formed by reversion and complementation followed by the total machines producing them.}
\begin{center}
\begin{tabular}{|r|r|}
\hline
 & \\
 \normalsize{string group} & \normalsize{\# occurrences} \\
\hline 0, 1 & 1224440064 \\
 01, 10 & 611436144 \\
 00, 11 & 611436144 \\
 001, 011, 100, 110 & 215534184 \\
 000, 111 & 112069020 \\
 010, 101 & 102247932 \\
 0001, 0111, 1000, 1110 & 23008080 \\
 0010, 0100, 1011, 1101 & 22675896 \\
 0000, 1111 & 14917104 \\
 0101, 1010 & 11425392 \\
 0110, 1001 & 9712752 \\
 0011, 1100 & 9628728 \\
 00001, 01111, 10000, 11110 & 2042268 \\
 00010, 01000, 10111, 11101 & 1984536 \\
 01001, 01101, 10010, 10110 & 1726704 \\
 00000, 11111 & 1683888 \\
 00110, 01100, 10011, 11001 & 1512888 \\
 00101, 01011, 10100, 11010 & 1478244 \\
 00011, 00111, 11000, 11100 & 1288908 \\
 00100, 11011 & 900768 \\
 01010, 10101 & 819924 \\
 01110, 10001 & 554304 \\
 000001, 011111, 100000, 111110 & 233064 \\
 000010, 010000, 101111, 111101 & 219552 \\
 000000, 111111 & 209436 \\
 010110, 011010, 100101, 101001 & 169896 \\
 001010, 010100, 101011, 110101 & 167964 \\
 000100, 001000, 110111, 111011 & 164520 \\
 001001, 011011, 100100, 110110 & 140280 \\
 010001, 011101, 100010, 101110 & 129972\\
 \hline
\end{tabular}
\end{center}
\label{default}
\end{table}
 
These are the top 10 string groups (i.e. with their reverted and complemented counterparts) appearing sooner than expected and getting away from their length blocks. That is, their lengths were greater than the next string in the classification order): 11111111, 11110111, 000000000, 111111111, 0000100
00, 111101111, 111111110, 010101010, 101010101, 000101010. This means these string groups  had greater algorithmic probability and therefore less algorithmic complexity than shorter strings.

Table \ref{stats} displays some statistical information of the distribution. The distribution is skewed to the right, the mass of the distribution is therefore concentrated on the left with a long right tail, as shown in Fig. 2.

\begin{table}[hdtp]
\label{stats}
\caption{Statistical values of the empirical distribution function $D(4)$ for strings of length $l=8$.}
\begin{center}
\begin{tabular}{c|c}
 & value\\
\hline
 \normalsize{mean} & 0.00391\\
 \normalsize{median} &  0.00280\\
 \normalsize{variance} & 0.0000136 \\
 \normalsize{kurtosis} & 23 \\
 \normalsize{skewness} & 3.6\\
  \hline
\end{tabular}
\end{center}
\label{default}
\end{table}

\begin{table}[htdp]
\caption{The probability of producing a string of length $l$ exponentially decreases as $l$ linearly increases. The  slowdown in the rate of decrease for string length $l>8$ is due to the few longer strings produced in $(4,2)$.}
\begin{center}
\begin{tabular}{|r|r|}
\hline
 \normalsize{length $n$} & \normalsize{$pr(n)$  } \\
 \hline
 1 & 0.410 \\
 2 & 0.410 \\
 3 & 0.144 \\
 4 & 0.0306 \\
 5 & 0.00469 \\
 6 & 0.000818 \\
 7 & 0.000110 \\
 8 & 0.0000226 \\
 9 & \normalsize{4.69$\times 10^{-6}$} \\
 10 & \normalsize{1.42$\times 10^{-6}$} \\
 11 & \normalsize{4.9$\times 10^{-7}$} \\
 12 & \normalsize{1.69$\times 10^{-7}$}\\
 \hline
\end{tabular}
\end{center}
\label{default}
\end{table}

\begin{figure}[htdp]
\label{stringlengthprobability}
\centering
   \scalebox{.8}{\includegraphics{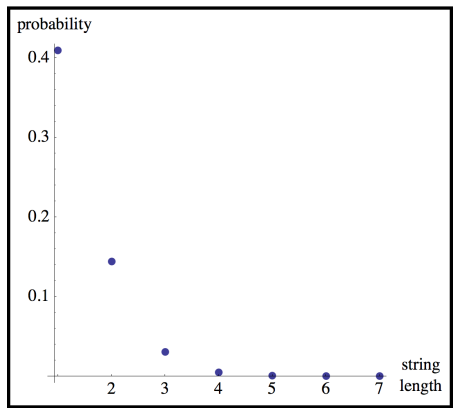}}
\caption{$(4,2)$ frequency distribution by string length.}
\end{figure}

\begin{figure}[htdp]
\label{dens}
\centering
   \scalebox{.62}{\includegraphics{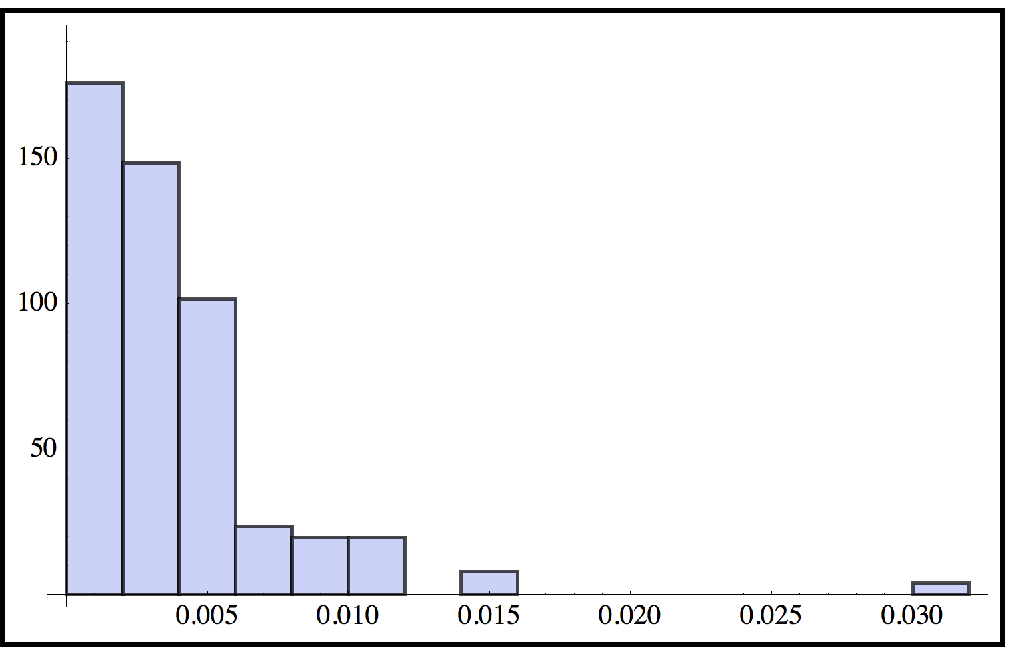}}
   \scalebox{.72}{\includegraphics{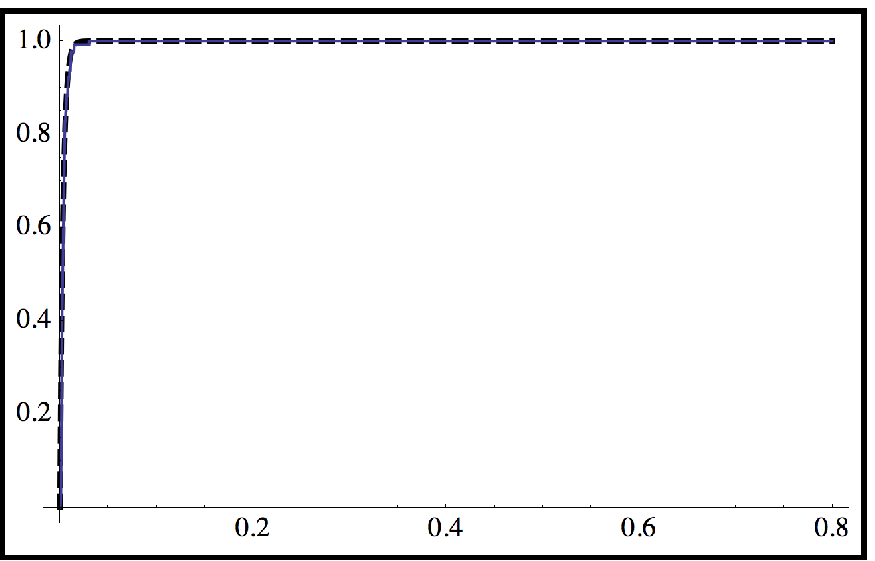}}  
\caption{Probability density function of bit strings of length $l=8$ from $(4,2)$. The histogram (left) shows the probabilities to fall within a particular region. The cumulative version (right) shows how well the distribution fits a Pareto distribution (dashed) with location parameter $k=10$. The reader may see but a single curve, that is because the lines overlap. $D(4)$ (and the sub-distributions it contains) is therefore log-normal.}
\end{figure}

\subsection{Derivation and calculation of the string's algorithmic complexity}

Algorithmic complexity values are calculated from the output probability distribution $D(4)$ through the application of the coding theorem and partially presented in Table \ref{complextable}. The full results are available online at \url{http://www.algorithmicnature.org} under the paper title as additional material.

\begin{table}[htdp]\renewcommand{\arraystretch}{.75}\addtolength{\tabcolsep}{-1pt}
\label{complextable}
\caption{Top 180 strings sorted from lowest to highest algorithmic complexity..}
\begin{center}
\begin{tabular}{|l|l|l|l|}
\hline
\footnotesize 0:2.29&\footnotesize 10110:12.76&\footnotesize 100100:16.38&\footnotesize 0100000:19.10\\
\footnotesize1:2.29&\footnotesize 01010:12.83&\footnotesize 110110:16.38&\footnotesize 1011111:19.10\\
\footnotesize00:3.29&\footnotesize 10101:12.83&\footnotesize 010001:16.49&\footnotesize 1111101:19.10\\
\footnotesize01:3.29&\footnotesize 00110:12.95&\footnotesize 011101:16.49&\footnotesize 0000100:19.38\\
\footnotesize10:3.29&\footnotesize 01100:12.95&\footnotesize 100010:16.49&\footnotesize 0010000:19.38\\
\footnotesize11:3.29&\footnotesize 10011:12.95&\footnotesize 101110:16.49&\footnotesize 1101111:19.38\\
\footnotesize000:5.74&\footnotesize 11001:12.95&\footnotesize 000011:16.49&\footnotesize 1111011:19.38\\
\footnotesize111:5.74&\footnotesize 00101:12.98&\footnotesize 001111:16.49&\footnotesize 0001000:19.45\\
\footnotesize001:5.79&\footnotesize 01011:12.98&\footnotesize 110000:16.49&\footnotesize 1110111:19.45\\
\footnotesize011:5.79&\footnotesize 10100:12.98&\footnotesize 111100:16.49&\footnotesize 0000110:19.64\\
\footnotesize100:5.79&\footnotesize 11010:12.98&\footnotesize 000110:16.52&\footnotesize 0110000:19.64\\
\footnotesize110:5.79&\footnotesize 00011:13.18&\footnotesize 011000:16.52&\footnotesize 1001111:19.64\\
\footnotesize010:5.87&\footnotesize 00111:13.18&\footnotesize 100111:16.52&\footnotesize 1111001:19.64\\
\footnotesize101:5.87&\footnotesize 11000:13.18&\footnotesize 111001:16.52&\footnotesize 0101110:19.68\\
\footnotesize0000:8.64&\footnotesize 11100:13.18&\footnotesize 001101:16.59&\footnotesize 0111010:19.68\\
\footnotesize1111:8.64&\footnotesize 01110:13.39&\footnotesize 010011:16.59&\footnotesize 1000101:19.68\\
\footnotesize0001:9.02&\footnotesize 10001:13.39&\footnotesize 101100:16.59&\footnotesize 1010001:19.68\\
\footnotesize0111:9.02&\footnotesize 000000:14.80&\footnotesize 110010:16.59&\footnotesize 0010001:20.04\\
\footnotesize1000:9.02&\footnotesize 111111:14.80&\footnotesize 001100:16.62&\footnotesize 0111011:20.04\\
\footnotesize1110:9.02&\footnotesize 000001:15.64&\footnotesize 110011:16.62&\footnotesize 1000100:20.04\\
\footnotesize0101:9.03&\footnotesize 011111:15.64&\footnotesize 011110:16.66&\footnotesize 1101110:20.04\\
\footnotesize1010:9.03&\footnotesize 100000:15.64&\footnotesize 100001:16.66&\footnotesize 0001001:20.09\\
\footnotesize0010:9.04&\footnotesize 111110:15.64&\footnotesize 011001:16.76&\footnotesize 0110111:20.09\\
\footnotesize0100:9.04&\footnotesize 000010:15.73&\footnotesize 100110:16.76&\footnotesize 1001000:20.09\\
\footnotesize1011:9.04&\footnotesize 010000:15.73&\footnotesize 000101:16.80&\footnotesize 1110110:20.09\\
\footnotesize1101:9.04&\footnotesize 101111:15.73&\footnotesize 010111:16.80&\footnotesize 0010010:20.11\\
\footnotesize0110:9.26&\footnotesize 111101:15.73&\footnotesize 101000:16.80&\footnotesize 0100100:20.11\\
\footnotesize1001:9.26&\footnotesize 010010:15.93&\footnotesize 111010:16.80&\footnotesize 1011011:20.11\\
\footnotesize0011:9.28&\footnotesize 101101:15.93&\footnotesize 001110:16.96&\footnotesize 1101101:20.11\\
\footnotesize1100:9.28&\footnotesize 010101:16.02&\footnotesize 011100:16.96&\footnotesize 0010101:20.15\\
\footnotesize00000:11.79&\footnotesize 101010:16.02&\footnotesize 100011:16.96&\footnotesize 0101011:20.15\\
\footnotesize11111:11.79&\footnotesize 010110:16.10&\footnotesize 110001:16.96&\footnotesize 1010100:20.15\\
\footnotesize00001:12.51&\footnotesize 011010:16.10&\footnotesize 001011:17.23&\footnotesize 1101010:20.15\\
\footnotesize01111:12.51&\footnotesize 100101:16.10&\footnotesize 110100:17.23&\footnotesize 0100101:20.16\\
\footnotesize10000:12.51&\footnotesize 101001:16.10&\footnotesize 000111:17.29&\footnotesize 0101101:20.16\\
\footnotesize11110:12.51&\footnotesize 001010:16.12&\footnotesize 111000:17.29&\footnotesize 1010010:20.16\\
\footnotesize00010:12.55&\footnotesize 010100:16.12&\footnotesize 0000000:18.03&\footnotesize 1011010:20.16\\
\footnotesize01000:12.55&\footnotesize 101011:16.12&\footnotesize 1111111:18.03&\footnotesize 0001010:20.22\\
\footnotesize10111:12.55&\footnotesize 110101:16.12&\footnotesize 0101010:18.68&\footnotesize 0101000:20.22\\
\footnotesize11101:12.55&\footnotesize 000100:16.15&\footnotesize 1010101:18.68&\footnotesize 1010111:20.22\\
\footnotesize00100:12.69&\footnotesize 001000:16.15&\footnotesize 0000001:18.92&\footnotesize 1110101:20.22\\
\footnotesize11011:12.69&\footnotesize 110111:16.15&\footnotesize 0111111:18.92&\footnotesize 0100001:20.26\\
\footnotesize01001:12.76&\footnotesize 111011:16.15&\footnotesize 1000000:18.92&\footnotesize 0111101:20.26\\
\footnotesize01101:12.76&\footnotesize 001001:16.38&\footnotesize 1111110:18.92&\footnotesize 1000010:20.26\\
\footnotesize10010:12.76&\footnotesize 011011:16.38&\footnotesize 0000010:19.10&\footnotesize 1011110:20.26\\
  \hline
\end{tabular}
\end{center}
\label{default}
\end{table}

The largest algorithmic complexity value after the application of the coding theorem was $\max{\{C(s)  :  s \in D(4)\}}=29$ bits. When interpreted as program size values it is worth mention that after application of the coding theorem the string frequencies obtained are often real numbers, one can either take the ceiling integer value or take it as a different (finer) measure closely related to algorithmic complexity, but not necessarily exactly the same (the Kolmogorov-Chaitin complexity is a norm, the Solomonoff-Levin complexity (algorithmic probability) is a frequency, the coding theorem says they converge in the limit).

\begin{figure}[htdp]
\label{frequencyprobability}
\centering
   \scalebox{.85}{\includegraphics{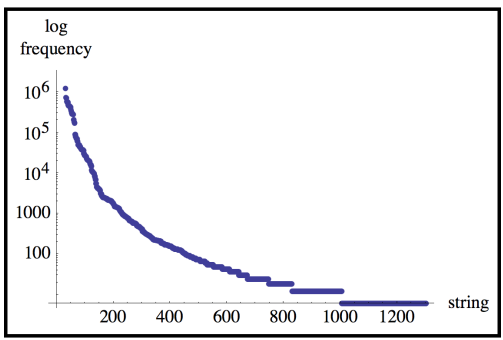}}
\caption{$(4,2)$ output log-frequency plot, ordered from most to less frequent string.}
\end{figure}

\subsubsection{Same length string complexity}

The complexity classification \ref{thetable} allows to make a comparison of the structure of the strings related to their calculated complexity among all the strings of the same length extracted from $D(4)$.

\begin{table}[hdtp]
\label{thetable}
\caption{Algorithmic complexity classification---from less to more random---for 7-bit strings extracted from $D(4)$ after application of the coding theorem.}
\begin{center}
\begin{tabular}{|c|c|c|c|}
\hline
0000000:18.03&1001000:20.09&0101001:20.42&0000111:20.99\\
1111111:18.03&1110110:20.09&0110101:20.42&0001111:20.99\\
0101010:18.68&0010010:20.11&1001010:20.42&1110000:20.99\\
1010101:18.68&0100100:20.11&1010110:20.42&1111000:20.99\\
0000001:18.92&1011011:20.11&0001100:20.48&0011110:21.00\\
0111111:18.92&1101101:20.11&0011000:20.48&0111100:21.00\\
1000000:18.92&0010101:20.15&1100111:20.48&1000011:21.00\\
1111110:18.92&0101011:20.15&1110011:20.48&1100001:21.00\\
0000010:19.10&1010100:20.15&0110110:20.55&0111110:21.03\\
0100000:19.10&1101010:20.15&1001001:20.55&1000001:21.03\\
1011111:19.10&0100101:20.16&0011010:20.63&0011001:21.06\\
1111101:19.10&0101101:20.16&0101100:20.63&0110011:21.06\\
0000100:19.38&1010010:20.16&1010011:20.63&1001100:21.06\\
0010000:19.38&1011010:20.16&1100101:20.63&1100110:21.06\\
1101111:19.38&0001010:20.22&0100010:20.68&0001110:21.08\\
1111011:19.38&0101000:20.22&1011101:20.68&0111000:21.08\\
0001000:19.45&1010111:20.22&0100110:20.77&1000111:21.08\\
1110111:19.45&1110101:20.22&0110010:20.77&1110001:21.08\\
0000110:19.64&0100001:20.26&1001101:20.77&0010011:21.10\\
0110000:19.64&0111101:20.26&1011001:20.77&0011011:21.10\\
1001111:19.64&1000010:20.26&0010110:20.81&1100100:21.10\\
1111001:19.64&1011110:20.26&0110100:20.81&1101100:21.10\\
0101110:19.68&0000101:20.29&1001011:20.81&0110001:21.13\\
0111010:19.68&0101111:20.29&1101001:20.81&0111001:21.13\\
1000101:19.68&1010000:20.29&0001101:20.87&1000110:21.13\\
1010001:19.68&1111010:20.29&0100111:20.87&1001110:21.13\\
0010001:20.04&0000011:20.38&1011000:20.87&0011100:21.19\\
0111011:20.04&0011111:20.38&1110010:20.87&1100011:21.19\\
1000100:20.04&1100000:20.38&0011101:20.93&0001011:21.57\\
1101110:20.04&1111100:20.38&0100011:20.93&0010111:21.57\\
0001001:20.09&0010100:20.39&1011100:20.93&1101000:21.57\\
0110111:20.09&1101011:20.39&1100010:20.93&1110100:21.57\\
\hline
\end{tabular}
\end{center}
\label{default}
\end{table}

\subsubsection{Halting summary}

\begin{figure}[htdp]
\label{stringlengthprobability}
\begin{center}
   \scalebox{.3}{\includegraphics{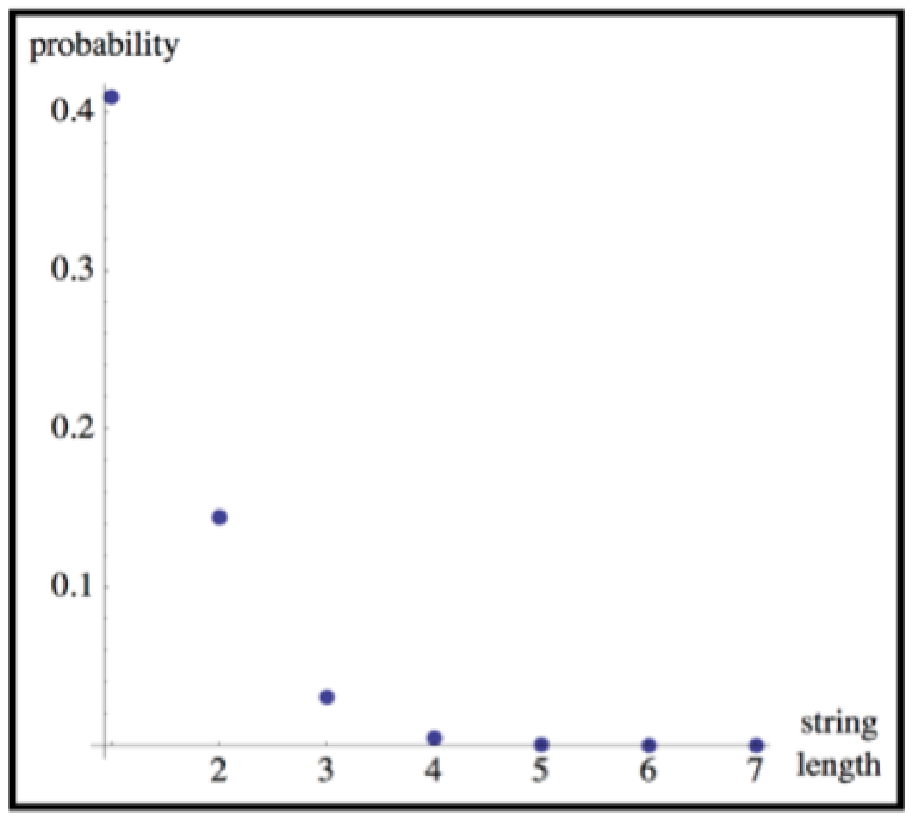}}
   \scalebox{.25}{\includegraphics{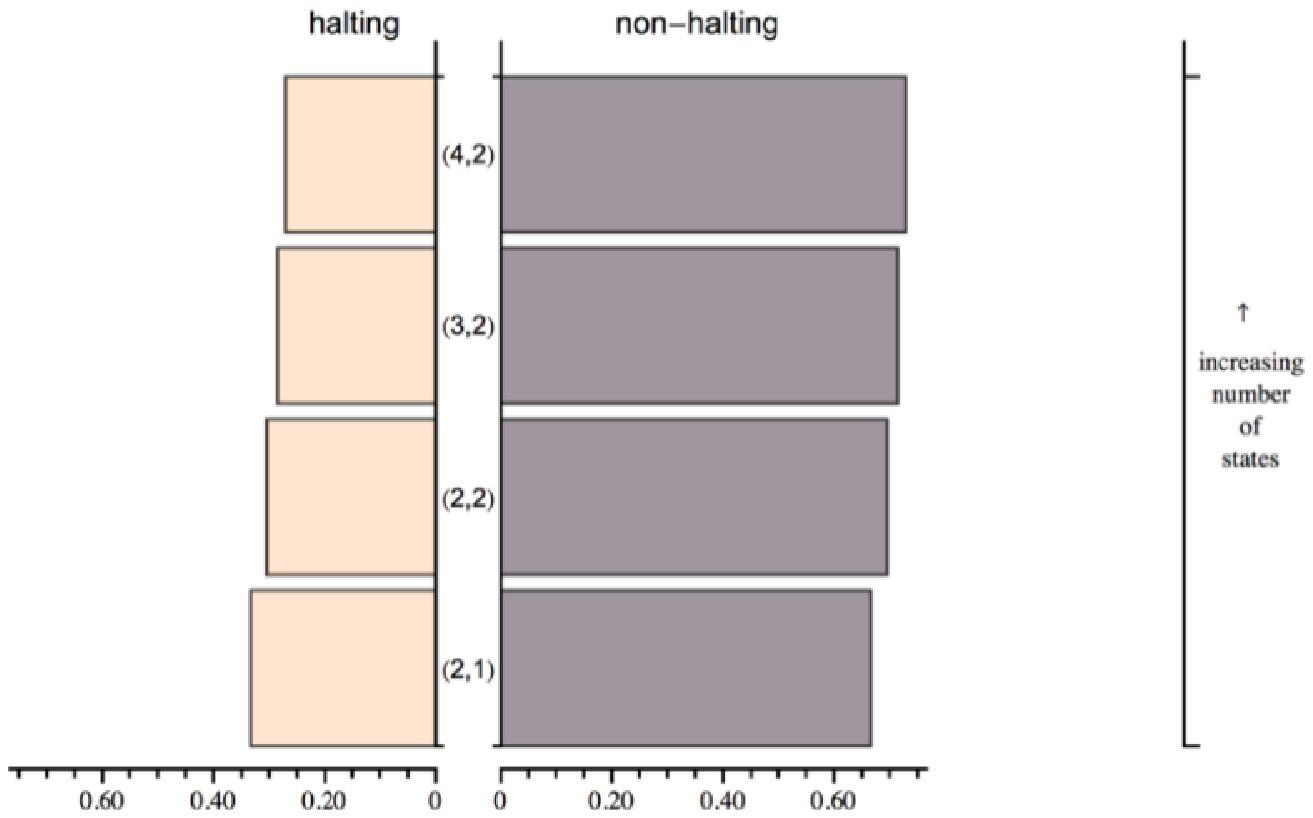}}
\end{center}
\caption{Graphs showing the halting probabilities among $(n,2)$, $n<5$. The list plot on the left shows the decreasing probability of the number of halting Turing machines while the paired bar chart on the right allows a visual comparison between both halting and non-halting machines side by side.}
\end{figure}

In summary, among the (running over a tape filled with 0 only): 12, 3\,044, 2\,147\,184 and 2\,985\,384\,480 Turing machines in $(n,2)$, $n<5$, there were 36, 10\,000, 7\,529\,536 and 11\,019\,960\,576 that halted, that is slightly decreasing fractions of 0.333..., 0.3044, 0.2851 and 0.2709 respectively.\\

\subsection{Runtimes investigation}

Runtimes much longer than the lengths of their respective halting programs are rare and the empirical distribution approaches the \emph{a priori} computable probability distribution on all possible runtimes predicted in  \cite{calude}. As reported in  \cite{calude} ``long'' runtimes are effectively rare. The longer it takes to halt, the less likely it is to stop.

\begin{figure}[h!]
\centering
   \scalebox{.65}{\includegraphics{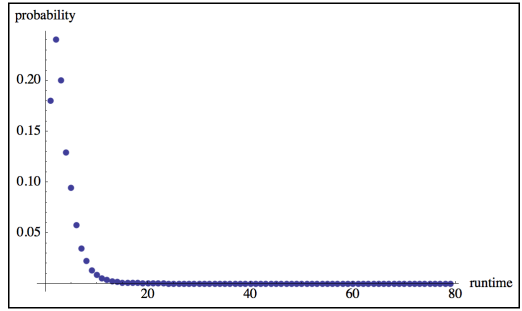}}
\caption{Runtimes distribution in $(4,2)$.}
\end{figure}

\begin{table}[hdtp]
\caption{Probability that a n-bit string among all $n<10$ bit strings is produced at
times $t<8$.}
\begin{center}
\begin{tabular}{c|cccccccc}
\hline
 & $t=1$ & $t=2$ & $t=3$ & $t=4$ & $t=5$ & $t=6$ & $t=7$\\
\hline
 \normalsize{n=1} & 1.0 & 0 & 0 & 0 & 0 & 0 & 0 & 0 \\
 \normalsize{n=2} & 0 & 1.0 & 0.60 & 0.45 & 0.21 & 0.11 & 0.052 & 0.025 \\
 \normalsize{n=3} & 0 & 0 & 0.40 & 0.46 & 0.64 & 0.57 & 0.50 & 0.36 \\
 \normalsize{n=4} & 0 & 0 & 0 & 0.092 & 0.15 & 0.29 & 0.39 & 0.45 \\
 \normalsize{n=5} & 0 & 0 & 0 & 0 & 0 & 0.034 & 0.055 & 0.16 \\
 \normalsize{n=6} & 0 & 0 & 0 & 0 & 0 & 0 & 0 & 0.0098 \\
 \normalsize{n=7} & 0 & 0 & 0 & 0 & 0 & 0 & 0 & 0 \\
 \normalsize{n=8} & 0 & 0 & 0 & 0 & 0 & 0 & 0 & 0 \\
 \normalsize{n=9} & 0 & 0 & 0 & 0 & 0 & 0 & 0 & 0 \\
 \normalsize{n=10} & 0 & 0 & 0 & 0 & 0 & 0 & 0 & 0 \\
 \hline
 \normalsize{Total} & 1 & 1 & 1 & 1 & 1 & 1 & 1 & 1\\
 \hline
\end{tabular}
\end{center}
\label{default}
\end{table}

\begin{table}[hdtp]
\caption{Probability that a n-bit string with $n<10$ is produced at time $t<7$.}
\begin{center}
\begin{tabular}{c|ccccccc|c}
\hline
 & $t=1$ & $t=2$ & $t=3$ & $t=4$ & $t=5$ & $t=6$ & $t=7$ & Total\\
\hline
 \normalsize{n=1} & 0.20 & 0 & 0 & 0 & 0 & 0 & 0 & 0.20 \\
 \normalsize{n=2} & 0 & 0.14 & 0.046 & 0.016 & 0.0045 & 0.0012 & 0.00029 & 0.20 \\
 \normalsize{n=3} & 0 & 0 & 0.030 & 0.017 & 0.014 & 0.0063 & 0.0028 & 0.070 \\
 \normalsize{n=4} & 0 & 0 & 0 & 0.0034 & 0.0032 & 0.0031 & 0.0022 & 0.012 \\
 \normalsize{n=5} & 0 & 0 & 0 & 0 & 0 & 0.00037 & 0.00031 & 0.00069 \\
 \normalsize{n=6} & 0 & 0 & 0 & 0 & 0 & 0 & 0 & 0 \\
 \normalsize{n=7} & 0 & 0 & 0 & 0 & 0 & 0 & 0 & 0 \\
 \normalsize{n=8} & 0 & 0 & 0 & 0 & 0 & 0 & 0 & 0 \\
 \normalsize{n=9} & 0 & 0 & 0 & 0 & 0 & 0 & 0 & 0 \\
 \normalsize{n=10} & 0 & 0 & 0 & 0 & 0 & 0 & 0 & 0 \\
\hline
 \normalsize{Total} & 0.21 & 0.14 & 0.076 & 0.037 & 0.021 & 0.011 & 0.0057\\
 \hline
\end{tabular}
\end{center}
\label{default}
\end{table}

\noindent Among the various miscellaneous facts from these results: 

\begin{myitemize}
\item All 1-bit strings were produced at $t=1$.
\item 2-bit strings were produced at all $2<t<14$ times.
\item $t=3$ was the time at which the first 2  bit strings of different lengths were produced ($n=2$ and $n=3$).
\item Strings produced before 8 steps account for 49\% of the strings produced by all $(4,2)$ halting machines.
\item There were 496 string groups produced by $(4,2)$, that is strings that are not symmetric under reversion or complementation.
\item There is a relation between $t$ and $n$; no n-bit string is produced before $t<n$. This is obvious because a machine needs at least $t$ steps to print $t$ symbols.
\item At every time $t$ there was at least one string of length $n$ for $1<n<t$.
\end{myitemize}

\section{Discussion}

Intuitively, one may be persuaded to assign a lower or higher algorithmic complexity to some strings when looking at tables 9 and 10, because they may seem simpler or more random  than others of the same length. We think that very short strings may appear to be more or less random but may be as hard to produce as others of the same length, because Turing machines producing them may require the same quantity of resources to print them out and halt as they would with others of the same (very short) length.

For example, is 0101 more or less complex than 0011? Is 001 more or less complex than 010? The string 010 may seem simpler than 001 to us because we may picture it as part of a larger sequence of alternating bits, forgetting that such is not the case and that 010 actually was the result of a machine that produced it when entering into the halting state, using this extra state to somehow delimit the length of the string. No satisfactory argument may exist to say whether 010 is really more or less random than 001, other than actually running the machines and looking at their objective ranking according to the formalism and method described herein. The situation changes for larger strings, when an alternating string may in effect strongly suggest that it should be less random than other strings because a short description is possible in terms of the simple alternation of bits. Some strings may also assume their correct rank when the calculation is taken further, for example if we were able to compute $D(5)$.

On the other hand,  it may seem odd that the program size complexity of a string of length $l$ is systematically larger than $l$ when $l$ can be produced by a \emph{print} function of length l+\{the length of the print program\}, and indeed one can interpret the results exactly in this way. The surplus can be interpreted as a constant product of a \emph{print} phenomenon which is particularly significant for short strings. But since it is a constant, one can subtract it from all the strings. For example, subtracting 1 from all values brings the complexity results for the shortest strings to exactly their size, which is what one would expect from the values for algorithmic complexity. On the other hand, subtracting the constant preserves the relative order, even if larger strings continue having algorithmic complexity values larger than their lengths. What we provide herein, besides the numerical values, is a hierarchical structure from which one can tell whether a string is of greater, lesser or equal algorithmic complexity.

The \emph{print program} assumes the implicit programming of the halting configuration. In C language, for example, this is delimited by the semicolon. The fact then that a single bit string requires a 2 bit ``program'' may be interpreted as the additional information represented by the length of the string; the fact that a string is of length $n$ is not the result of an arbitrary decision but it is encoded in the producing machine. In other words, the string  not only carries the information of its $n$ bits, but also of the delimitation of its length. This is different to, for example, approaching the algorithmic complexity by means of cellular automata--there being no encoded halting state, one has to manually stop the computation upon producing a string of a certain arbitrary length according to an arbitrary stopping time. This is a research program that we have explored before \cite{zenildelahaye} and that we may analyze in further detail somewhere else.

It is important to point out that after the application of the coding theorem one often gets a non-integer value when calculating $C(s)$ from $m(s)$. Even though when interpreted as the size in bits of the program produced by a Turing machine it should be an integer value because the size of a program can only be given in an integer number of bits. The non-integer values are, however, useful to provide a finer structure providing information on the exact places in which strings have been ranked.

An open question is how much of the relative string order (hence the relative algorithmic probability and the relative algorithmic complexity) of $D(n)$ will be preserved when calculating $D(i)$ for larger Turing machine spaces such that $0<n<i$. As reported here, $D(n)$ preserves most of the string orders of $D(n-1)$ for $1<n<5$. While each space $(n,2)$ contains all $(n-1,2)$ machines, the exponential increase in number of machines when adding states may easily produce strings such that the order of the previous distribution is changed. What the results presented here show, however, is that each new space of larger machines contributes in the same proportion to the number of strings produced in the smaller spaces, in such a way that they preserve much of the previous string order of the distributions of smaller spaces, as shown by calculating the Spearman coefficient indicating a very strong ranking correlation. In fact, some of the ranking variability between the distributions of spaces of machines with different numbers of states occurred later in the classification, likely due to the fact that the smaller spaces missed the production of some strings. For example, the first rank difference between $D(3)$ and $D(4)$ occurred in place 20, meaning that the string order in $D(3)$ was strictly preserved in $D(4)$ up to the top 20 strings sorted from higher to lower frequency. Moreover, one may ask whether the actual frequency values of the strings converge.

\section{Concluding remarks}
\label{conclusions}

We have provided numerical tables with values the algorithmic complexity for short strings, and we have shed light into the behavior of small Turing machines, particularly halting runtimes and output frequency distributions. The calculation of $D(n)$ provides an empirical and \emph{natural} distribution that does not depend on an additive constant and may be used in several practical contexts. The approach, by way of algorithmic probability, also reduces the impact of the additive constant given that one does not seem to be forced to make many arbitrary choices other than fixing a standard model of computation (as opposed to fixing a specific universal Turing machine). In other words, the approach is bottom-up rather than top-down.

An interesting open question is how robust the produced complexity classifications are to variations in the computational description formalism, such as using Turing machines with one-directional tapes rather than bi-directional, or following completely different models such as $n$-dimensional cellular automata, or Post tag systems. We've shown in \cite{zenildelahaye} that reasonable formalisms seem to produce reasonable complexity classifications, in the sense that: a) they are close to what intuition would tell should be and b) they are statistically correlated with each other at various degrees of confidence. This is, however, a topic of current investigation.

\section*{Acknowledgment}

Hector Zenil wants to thank Matthew Szudzik for his always valuable advice.

\bibliographystyle{elsarticle-num}

\end{document}